\documentclass[a4paper,12pt]{article}

\usepackage[pdftex]{graphicx}
\usepackage[english]{babel}
\usepackage{amsmath,amssymb}
\usepackage{relsize} 

\author{L.J. Lehman$^{1*}$, V. Zatloukal$^2$, J.K. Pachos$^3$, G.K. Brennen$^1$}
\title{Braiding Interactions in Anyonic Quantum Walks}
\date{}

\graphicspath{{./figs/}}

\newcommand{\bra}[1]{\big\langle#1\big\vert}
\newcommand{\ket}[1]{\big\vert#1\big\rangle}
\newcommand{\tr}{\text{Tr}}

\begin{document}

\maketitle

\begin{center}
$^1$ Centre for Engineered Quantum Systems (EQuS), Macquarie University, \\NSW 2109, Australia\\
$^2$ Faculty of Nuclear Sciences and Physical Engineering, Czech Technical University\\ in Prague,
B\v{r}ehov\'{a} 7, 115 19 Praha 1, Czech Republic\\
$^3$ School of Physics and Astronomy, University of Leeds, Leeds LS2 9JT, UK\\
$^*$ email: lauri.lehman@students.mq.edu.au
\end{center}

\begin{abstract}
The anyonic quantum walk is a dynamical model describing a single anyon propagating along
a chain of stationary anyons and interacting via mutual braiding statistics.
We review the recent results on the effects of braiding statistics in anyonic quantum walks in quasi-one dimensional ladder geometries.  For anyons which correspond to spin-$1/2$ irreps of the quantum groups $SU(2)_k$, the non-Abelian species $(1<k<\infty)$ gives rise to entanglement between the walker and topological degrees of freedom which is quantified by quantum link invariants over the trajectories of the walk.  The decoherence is strong enough to reduce the walk on the infinite ladder to classical like behaviour.  We also present numerical results on mixing times of $SU(2)_2$ or Ising model anyon walks on cyclic graphs.  Finally, the possible experimental simulation of the anyonic quantum walk in Fractional Quantum Hall systems is discussed.
\end{abstract}

\section{Introduction}

In two spatial dimensions, the exchange statistics of identical particles must be generalized to
braiding statistics \cite{lm1977}. The particles which obey braiding statistics are called anyons,
and they are thought to exist as quasiparticle excitations in Fractional Quantum Hall systems \cite{nssfs2008} or as
edge modes of semiconductor nanowires \cite{aorof2011}. Much of the interest in anyons is motivated by the
realization that particles with non-Abelian braiding statistics can be used to perform quantum computation
in a decoherence-free manner \cite{pachos2012}. In such schemes the information is encoded in the collective
many-body state of a system in a topological phase of matter.

Information stored in the collective state of anyons can be manipulated by either fusing or braiding anyons.
Recently there has been some research activity on the equilibrium properties of systems of anyons
interacting via fusion \cite{ftltkwf2007,tafhlt2008,lptt2011}. Here we discuss the non-equilibrium properties
of anyons which interact via braiding. The dynamics of anyons can be simulated by a discrete-time quantum
walk, where the motion of a moving anyon is coupled to an external degree of freedom (DOF) called the quantum coin.
The moving anyon, called the walker, braids with an array of stationary anyons arranged on a line.
We concentrate on the effects of braiding on the propagation of the moving anyon along the lattice.

If the anyons are non-Abelian, the mobile anyon and the stationary anyons possess a collective Hilbert space
$\mathcal{H}_{\text{F}}$, which grows exponentially in the number of anyons, and which allows for interactions between them.  
In fact this coupling induces decoherence on spatial DOF of the mobile anyon, and we study the effects of this coupling
in both finite and infinite lattices. Differences between Abelian and non-Abelian anyons are pointed out,
and possible implications for distinguishing between these two types of anyons experimentally are discussed.

Phase factors in two-particle walks, corresponding to braiding phases of Abelian anyons
were studied by Berry and Wang \cite{bw2011} and a similar scheme was implemented experimentally with integrated
photonics by Sansoni {\it et al} \cite{ssvmcro2012}.  Also there has been some related work on using quantum walks to simulate different phases of topological insulators
\cite{krbd2010,kbfrbkadw2012,kitagawa2012,ok2011}.  In such systems, a discrete walk models coarse grained single particle continuous dynamics of a system possessing bound states at the boundaries of different symmetry protected topological phases.  These bound states are robust to small perturbations of the system which respect the symmetry (e.g. time reversal symmetry) and can be probed by varying the step dynamics in the quantum walk.  In contrast our model involves multiple particles with topological symmetry which is robust to arbitrary local perturbations.

\section{Anyonic quantum walk}

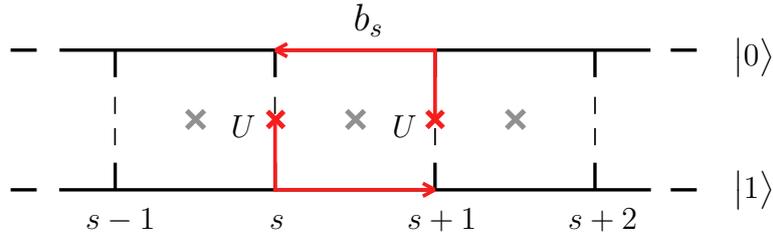
\begin{figure}
\begin{center}
\setlength{\unitlength}{.6\textwidth}
\begin{picture}(1,.32)
\put(0,.04){\includegraphics[width=.9\unitlength]{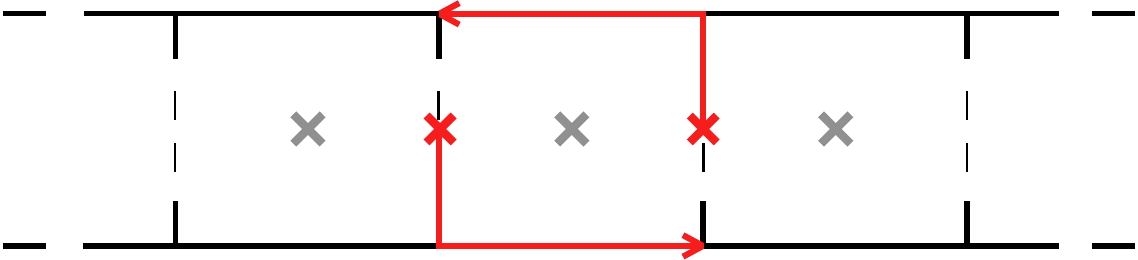}}
\put(.945,.215){$\ket{0}$}
\put(.945,.04){$\ket{1}$}
\put(.29,.12){$U$}
\put(.5,.12){$U$}
\put(.1,0){$s-1$}
\put(.34,0){$s$}
\put(.52,0){$s+1$}
\put(.73,0){$s+2$}
\put(.45,.26){\larger[1]$\mathlarger{b_s}$}
\end{picture}
\end{center}
\caption{The anyonic quantum walk model. Here the coin modes $0$ and $1$ label the position of the walker
(red cross) on either upper or lower edge, respectively. The locations of the contacts between the edges are
labelled by spatial sites $s$. The coin flip matrix $U$ corresponds to a scattering matrix at the edge contacts.
The action of braiding between the walker anyon and the stationary anyon (grey cross) between sites $s$
and $s+1$ is represented by the generator $b_s$.}
\label{fig:aqw}
\end{figure}

The dynamical behaviour of anyons can be studied by decimating the time evolution into discrete steps
of infinitesimal length. Here we consider a single particle (walker) hopping on a one-dimensional lattice of
spatial sites. Between each discrete time step, the walker is allowed to move only to its neighbouring
sites on the left and right. To study the effects of anyonic interactions in the dynamics, we place an
array of stationary anyons on the dual lattice between the spatial sites as shown in Fig. \ref{fig:aqw}.
The mobile anyon and the stationary anyons are assumed to stay far enough apart from each other such that the
interactions are mediated exclusively by braiding statistics. Although the spatial lattice
extends in one dimension only, it is necessary to think of the system as two-dimensional, such that the
mobile anyon can circumnavigate the stationary anyons without coming into contact with them.

The braiding interactions of anyons are purely topological. The action on the wave function does not depend on
the exact path of the walker, only on the initial and final points \footnote{There could be dynamical and or geometric phases arising from e.g. Aharanov-Bohm interactions but we neglect them in the following.  Provided the interactions that give rise to these phases are quenched in time, they do not qualitatively  affect the transport behaviour of non-Abelian anyons.  If they are random in space they can induce localization in Abelian anyons \cite{zlspb2012}.}.  For simplicity of analysis we consider only counterclockwise braids, meaning the system is \emph{chiral}. Numerical results on continuous-time anyonic walks show that
the essential features of the walk do not change for non-chiral walks \cite{zlspb2012,spvb}.

The formalism of anyonic quantum walks \cite{bekptw2010} is a generalization of the discrete-time Hadamard walk on an infinite
one-dimensional lattice as defined by Ambainis \textit{et al} \cite{abnvw2001}.
In the anyonic quantum walk model the Hadamard walk is changed to include the effects of the
braiding interactions.
In addition to the spatial and coin degrees of freedom, one has to accommodate for the
fusion Hilbert space of (non-Abelian) anyons, and the total Hilbert space is thus
$\mathcal{H}=\mathcal{H}_{\text{S}}\otimes\mathcal{H}_{\text{F}}\otimes\mathcal{H}_{\text{C}}$.
The space $\mathcal{H}_{\text{S}}$  of $N$ spatial sites is spanned by the basis vectors $\{\ket{s}\}_{s=1}^N$, $\mathcal{H}_{\text F}$ is described below, and the two-dimensional
coin space $\mathcal{H}_{\text{C}}$ is spanned by $\{\ket{0},\ket{1}\}$, where $0$ corresponds to going left and $1$ to going right.  
The system is initially in the product state
$\ket{\Psi(0)}=\ket{s_0}_{\text{S}}\ket{\Phi_0}_{\text{F}}\ket{c_0}_{\text{C}}$.
In the initial configuration, the walker is localized at some initial
site $s_0$ and the coin state is chosen to be $\ket{c_0}=\ket{0}$. The initial state of the fusion space $\ket{\Phi_0}$ is the
vacuum state such that the anyons taking part in the walk are created in particle-antiparticle pairs.
If the particle-antiparticle pairs reside on neighboring islands, the tracing scheme is determined to
be the plat closure. If the antiparticles are thought to be ancillary particles which are dragged out of the system
for the duration of the walk, the tracing scheme is Markov.

The time evolution between infinitesimal time steps is given by the operator $\ket{\Psi(t+1)}=W\ket{\Psi(t)}$
which consists of the coin flip operator and a subsequent conditional braiding operator {$W$}{$=$}{$TU$}.
The coin flip operator acts like the Hadamard gate on the coin DOF:
$U=\mathbb{I}_{\text{S}}\otimes\mathbb{I}_{\text{F}}\otimes\frac{1}{\sqrt{2}}\begin{pmatrix}1&1\\1&-1\end{pmatrix}$,
and the conditional braiding operator moves the walker to the adjacent lattice sites, conditioned on the coin
state and simultaneously braiding with the stationary anyon between the lattice sites (see Fig. \ref{fig:aqw}):
\begin{equation}
T\big(\ket{s}\ket{f}\ket{c}\big) = \ket{s-1}b_{s-1}\ket{f}\ket{0}\langle0\ket{c}
+ \ket{s+1}b_s\ket{f}\ket{1}\langle1\ket{c}.
\end{equation}
The wave function at time $t$ is obtained by applying the single step operator repeatedly on the initial state
$\ket{\Psi(t)}=W^t\ket{\Psi(0)}$.

\subsection{Representations of the braid group}

The generators of the braid group $b_s\in B_{N_A}$ act in the fusion space $\mathcal{H}_{\text{F}}$ only.
The space $\mathcal{H}_{\text{F}}$ is spanned by vectors $\{\ket{g_1,g_2,\ldots}\}$ where the quantum numbers
$g_j$ label possible intermediate fusion outcomes of $N_A$ identical anyons. The representations of $b_s$
depend on the \emph{anyon model}, which is defined by the set of all charge types and the fusion rules of the
charges. Here we give the representations of $b_s$ for two different particle types, the $\sigma$ charges of
the so called Ising anyon model, and the spin-$1/2$ irreducible representations (irreps) of the quantum group $SU(2)_2$.
The fusion spaces of these particles have a very convenient tensor product structure in terms of qubits.
Every even intermediate fusion charge of an $N_A$-anyon state in the standard basis is constrained to be
{$g_{2j}$}{$=$}{$\sigma$} (or {$g_{2j}$}{$=$}{$\frac{1}{2}$}) for the Ising and $SU(2)_2$ models respectively,
and every odd charge is constrained to two values {$g_{2j+1}$}{$\in$}{$\{1,\psi\}$} (or {$g_{2j+1}$}{$\in$}{$\{0,1\}$}),
forming a single qubit.
Requiring that the total charge is vacuum, the fusion space then consists of $m=N_A/2-1$ qubits,
$\mathcal{H}_{\text{F}}\simeq(\mathbb{C}^2)^{\otimes m}$.
The braid generators of Ising and $SU(2)_2$ anyons thus admit a simple structure where the non-trivial
matrices acting on the fusion space are at most 4-by-4 \cite{bekptw2010}.
The even and odd generators can be expressed as
\begin{align}
&b_1=R\otimes_{j=2}^m\mathbb{I}_2,\quad
b_2=B\otimes_{j=2}^m\mathbb{I}_2,\quad
b_3=P\otimes_{j=3}^m,\notag\\
&b_{2k}=\otimes_{j=1}^{k-1}\mathbb{I}_2\otimes B\otimes_{j=k+1}^m\mathbb{I}_2,\quad
b_{2k+1}=\otimes_{j=1}^{k-1}\mathbb{I}_2\otimes P\otimes_{j=k+2}^m\mathbb{I}_2;\quad 1<k<m,\label{eq:braidgens}\\
&b_{N_A-3}=\otimes_{j=1}^{m-2}\mathbb{I}_2\otimes P,\quad
b_{N_A-2}=\otimes_{j=1}^{m-1}\mathbb{I}_2\otimes B,\quad
b_{N_A-1}=\otimes_{j=1}^{m-1}\mathbb{I}_2\otimes R.\notag
\end{align}
For the $\sigma$ anyons in the Ising model
the non-trivial matrices are
\begin{equation} \label{eq:isingmatrices1}
R=R_{\sigma,\sigma}=e^{-i\frac{\pi}{8}}\begin{pmatrix}1&0\\0&i\end{pmatrix},\quad
B=\frac{e^{-i\frac{\pi}{8}}}{\sqrt{2}}\begin{pmatrix}e^{i\frac{\pi}{4}}&e^{-i\frac{\pi}{4}}\\
e^{-i\frac{\pi}{4}}&e^{i\frac{\pi}{4}}\end{pmatrix}, \quad P=e^{-i\frac{\pi}{8}}{\rm diag}(1,i,i,1).
\end{equation}

The $R$ symbols of the $SU(2)_2$ model, describing the braiding of two particles, are obtained
from those of the Ising model by the substitution $R_{\frac{1}{2},\frac{1}{2}}^*=-iR_{\sigma,\sigma}$
where the star denotes complex conjugation.
The $F$ symbols, which describe changing the order of fusion for three particles,
are identical for the two models, explicitly $F=F_{\frac{1}{2}\frac{1}{2}\frac{1}{2}}^{\frac{1}{2}}=\frac{1}{\sqrt{2}}\begin{pmatrix}1&1\\1&-1\end{pmatrix}$.
Thus up to phase factors and complex conjugation, the Ising and
$SU(2)_2$ models act similarly under braiding of spin-$1/2$ irreps with charge correspondence
$\{1,\sigma,\psi\}\:\widehat{=}\:\{0,\frac{1}{2},1\}$.
Consequently the braid generators for the $SU(2)_2$ anyons are given by Eq. (\ref{eq:braidgens})
and substituting the matrices in Eq. (\ref{eq:isingmatrices1}) with
$R\rightarrow iR^*$, $B\rightarrow iB^*$ and $P\rightarrow iP^*$.
In most cases the difference between these two models is not distinguishable.

\subsection{Probability distribution and quantum link invariants} \label{sec:}

The main interest in the anyonic quantum walk is in how the occupation probability of the mobile anyon on the lattice
sites evolves, if it is initially localized at some initial site $s_0$.
The probability to be at site $s$ at time step $t$ is given by the diagonal values of the reduced density matrix:
\begin{equation} \label{eq:aqwprobdistr1}
p(s,t) = \big(\rho_{\text{S}}(t)\big)_{s,s}
= \bra{s}\tr_{\text{F}}\tr_{\text{C}}\big(W^t\ket{\Psi(0)}\bra{\Psi(0)}(W^\dag)^t\big)\ket{s}.
\end{equation}
In a $t$-step walk, the total transition amplitude between sites $s_0$ and $s$ can be given
as a sum over all the possible ways to end up on site $s$ after taking $t$ steps:
\begin{equation} \label{eq:aqwprobdistr2}
p(s,t) = \frac{1}{2^t}\sum_{\vec{a},\vec{a}'\leadsto s}(-1)^{z(\vec{a})+z(\vec{a}')}\;
\tr\:\mathcal{Y}(\vec{a},\vec{a}',\Phi_0)
\end{equation}
where paths are labelled by vectors $\vec{a}=\{a_1,a_2,\ldots,a_t\}$ with {$a_j$}{$=$}{$0$} ({$a_j$}{$=$}{$1$})
if the walker went left (right) at time step $j$.
The quantity $z(\vec{a})=\sum_{j=1}^{t-1}a_ja_{j+1}$ counts the number of subsequent right-moves in the path $\vec{a}$,
and the anyonic term is given by
$\mathcal{Y}(\vec{a},\vec{a}',\Phi_0) = \mathcal{B}_{\vec{a}}\ket{\Phi_0}\bra{\Phi_0}\mathcal{B}_{\vec{a}'}^\dag$
with the \emph{braid word} $\mathcal{B}_{\vec{a}}$ given as a product of generators corresponding to path $\vec{a}$:  $\mathcal{B}_{\vec{a}}=\prod_{r=0}^{t-1}b_{s_0+2(\sum_{j=1}^{t-r}a_j)-(t-r)}$.
The notation $\vec{a},\vec{a}'\leadsto s$ means that the summation is done only over paths which satisfy
$\sum_{j=1}^ta_j = \sum_{j=1}^ta_j' = \frac{s-s_0+t}{2}$ and $a_t = a_t'$, constraints which arise from taking the diagonal elements of the spatial DOF and tracing over the coin.
The trace over the anyonic DOFs for spin-$1/2$ irreps of $SU(2)_k$
can be expressed in terms of link invariant polynomials,
as shown by Aharonov {\it et al} \cite{ajl2006}:
\begin{align} \label{eq:fusiontracejones}
\tr\:\mathcal{Y}(\vec{a},\vec{a}',\Phi_0)
= \bra{\Phi_0}\mathcal{B}_{\overline{a}'}^\dag\mathcal{B}_{\overline{a}}\ket{\Phi_0}
&= \frac{(-q^{3/4})^{w(L(\overline{a},\overline{a}'))}}{d^{N_A-1}}
V_{L(\overline{a},\overline{a}')}(q)\\
&=\frac{\big\langle L(\overline{a},\overline{a}')\big\rangle(A)}{d^{N_A-1}}
\end{align}
where $V_L(q)$ is the Jones polynomial of a link $L$ with parameter $q=e^{i\frac{2\pi}{k+2}}$,
$\big\langle L(\overline{a},\overline{a}')\big\rangle(A)$ is the Kauffman bracket with parameter
$A=q^{-1/4}=e^{\frac{-i\pi}{2(k+2)}}$, and the quantum dimension is
$d=-A^2-A^{-2}=2\cos\frac{\pi}{k+2}$. 
By choosing the initial state $\Phi_0$ suitably, namely as a product of vacuum pairs with one member of each pair taking part in braiding, the link $L(\overline{a},\overline{a}')$
is the Markov closure of the braid word $\mathcal{B}_{\overline{a}'}^\dag\mathcal{B}_{\overline{a}}$.

\subsection{Infinite chains} \label{sec:infinitechains}

When there are no boundaries in the system, the walker can extend to infinite lengths on the left and right.
If the walker is initially confined in some finite region, locality of the shift operators implies that
there is a maximum distance that the walker can propagate in a finite time $t$. The main question is
that if the walker is initially localized at some initial site $s_0$, what is the expected distance of
the walker from $s_0$ at time step $t$? The square of the expected distance is quantified by the variance
\begin{equation}
\sigma^2(t) = \big\langle(s-s_0)^2\big\rangle = \sum_sp(s,t)s^2-2s_0\sum_sp(s,t)s+s_0^2.
\end{equation}
One can identify different qualitative behaviour of walks by the functional dependence of the variance on the
number of time steps $t$. If $\sigma^2(t)=K_2t^2+K_1t+K_0$ the mean velocity of the walker is linearly
proportional to the maximum allowed velocity, and the behaviour is called ballistic.
This is the behaviour of the Hadamard walk without decoherence. Random walks and quantum walks
with decoherence scale as
$\sigma^2(t)=K_1t+K_0$ and propagation is called diffusive. Localization implies that $\sigma^2(t)<K_0$,
but the exact definition of localization can vary \cite{lehman2012}.

For Abelian anyons, the walker acquires a global phase $e^{i\varphi/2}$ from each braid with the
stationary anyons. The total phases for bra- and ket-evolution of the density matrix are $e^{it\varphi/2}$
and $e^{-it\varphi/2}$ respectively, so that the total effect is trivial and the braiding statistics
plays no role in the time evolution.
Non-trivial effects do arise for Abelian anyons if one introduces clockwise braiding and extends
the coin space to four dimensions. In this case the behaviour is ballistic for all values of the exchange
angle $\varphi$, but the coefficient of the leading term depends on $\varphi$ \cite{bekptw2010}.

The anyonic quantum walk of $SU(2)_2$ anyons was analysed in Ref. \cite{lzbpw2011}.
For the value of the parameter $q=i$, the Jones polynomial can be evaluated efficiently and
the probability distribution can be written as
\begin{equation} \label{eq:probdistraqw3}
p(s,t)=\sum_{
\overline{a},\overline{a}'\leadsto s}
\left\{\begin{array}{ll}(-1)^{z(\overline{a})+z(\overline{a}')+\tau(\overline{a},\overline{a}')}/2^t&L\text{ proper}\\
0&L\text{ not proper}\end{array}\right.
\end{equation}
where properness and $\tau(\overline{a},\overline{a}')$ are properties of the link corresponding to the path
$(\overline{a},\overline{a}')$. Numerical results show that the walk is diffusive, and the variance
was shown to be linearly dependent on $t$ at the asymptotic limit $t\rightarrow\infty$.
These results hold equivalently for $SU(2)_2$ and Ising anyons.

The classical diffusion like behaviour of the non-Abelian Ising anyons is not special to that model but in fact is generic for a large class of non-Abelian anyons.  To show this we investigated quantum walk behaviour of spin$-1/2$ irreps of all $SU(2)_k$ models.  For $k>2$ and finite these models have non-Abelian anyons.    An obstacle to the analysis, however, is that for higher values of $k$, the evaluation of the Jones polynomial is exponentially hard in the
number of time steps, and the algorithm that was used for {$k$}{$=$}{$2$} does not work.
Numerical results for {$t$}{$=$}{$10$} show that the distribution looks classical for small values of
$k$ and approaches quantum distribution when $k$ increases \cite{bekptw2010}.
Results for higher number of time steps can be obtained by adopting the so called $W^2$ model \cite{ldb2011}
where the walker evolves according to the time evolution operator $W$ for two time steps, after
which the fusion and coin DOF are traced out. A single step of the walk is described by the
completely positive trace preserving map
\begin{equation}
\rho_{\text{S}}(t+1) = \mathcal{E}(\rho_{\text{S}}(t))
= \sum\limits_{f,c}E_{fc}\;\rho_{\text{S}}(t)\;E_{fc}^\dag
\end{equation}
where $E_{fc}$ are Kraus generators, labelled by the basis states of the fusion and coin DOF.
Numerical results show that the walk is also diffusive for all values of $k$.
Approximative expression for $SU(2)_2$/Ising anyons for large $t$ can be obtained by approximating the Kraus
generators as circulant matrices. The result is
\begin{equation}
p(s,t) = \frac{1}{2^t}\binom{t}{\frac{2t-(s-s_0)}{4}}
\end{equation}
which is the probability distribution of the classical random walk.

The behaviour of non-Abelian walks can be explained by decoherence effects in quantum systems.
The spatial and coin DOFs become entangled with fusion DOF, and the
state is not pure anymore in the space $\mathcal{H}_{\text{S}}\otimes\mathcal{H}_{\text{C}}$.
This can be seen by considering the representations of the generators of the braid group as
given in Eq. (\ref{eq:braidgens}). When the walker propagates along
the lattice, it braids with a growing number of stationary anyons, and the fusion qubits are multiplied by matrices $R$, $B$ and $P$.
Consider the state of two previously untangled fusion qubits $i$ and $j$:
$\ket{\psi_1}=\big(\alpha\ket{0}+\beta\ket{1}\big)_i\big(\alpha\ket{0}+\beta\ket{1}\big)_{j}$, with $\alpha=(F_{\frac{1}{2}\frac{1}{2}\frac{1}{2}}^{\frac{1}{2}})_0^0$ and
$\beta=(F_{\frac{1}{2}\frac{1}{2}\frac{1}{2}}^{\frac{1}{2}})_1^0$.
These could represent one encoded qubit at the current position of the walker on the fusion tree and the other among adjacent anyons not yet braided around by the walker,
represented in a basis where the two fusion qubits belong to separated fusion trees with three charges each and total charge for all six equal to zero  \cite{bips2009}.  In the sector of $\frac{1}{2}$ total charge for each tree, the single intermediate charge of each tree defines a qubit. 
The walker becomes linked nontrivially with a stationary anyon if it braids twice with the stationary anyon.
This operation is represented by the square of their mutual braid generator, $\ket{\psi_2}=b^2\ket{\psi_1}$.
The amount of entanglement between the fusion qubits generated by the braid $b^2$ can be quantified by the linear entropy of the reduced density matrix
of the qubit $i$: $Q(\rho_{i})=2(1-\tr(\rho^2_{i})/\tr(\rho_{i})^2)$ where $\rho_{i}=\tr_{j}[P_{i,j}\ket{\psi_2}\bra{\psi_2}]$.  Here $P_{i,j}$ is a projector onto the two qubit subspace, necessary since the braiding can couple to a fifth leakage state with total charge in each tree equal to $\frac{3}{2}$, and the linear entropy normalises by the trace of the reduced state $\rho_{i}$.
We find that the linear entropy is maximal $Q(\rho_{i})=0.938$ for the level $k=3$ of Chern-Simons theory, and decreases as a function of
$k$, with an asymptotic behaviour $Q(\rho_{i})=\frac{225 \pi^2}{64k^2}-O(1/k^3)$. This is consistent with the fact that the representations of $SU(2)_k$ approach those of $SU(2)$
at the limit of large $k$, and the the action of two-fold exchange is trivial in $SU(2)$ (no entanglement generated).
Furthermore, the linear entropy is zero for $k=1$, which corresponds to Abelian anyons by fusion rules.
The fusion space dimension of Abelian anyons is one and no entanglement is therefore generated by braiding.

All the results discussed here hold for uniform configurations of anyons,
ie. there is exactly one anyon with multiplicity 1 between each spatial site.
More generally, one could define that each island between the lattice sites can host more
than one anyon. The transport properties of anyons in such
random charge configurations were considered in Ref. \cite{zlspb2012} with dynamical localization
phenomena in mind. This might be relevant for experimental simulations of anyonic quantum walks
since the quasiparticle occupations might be hard to control in experimental conditions.

\section{Finite chains} \label{sec:finitechains}

Finite lattices can be modelled by truncating the graph on the left and right and introducing boundaries at
these locations. This requires a choice of boundary conditions for the time evolution operator $W$.
Possible choices for unitary walks are periodic and reflective boundary conditions, and for non-unitary
walks, absorbing boundary conditions. More generally, one could think of the boundary as an interface between
the quantum system and its environment, such that the walker interacts with the environment at the boundary,
and this interaction changes the state of the walker in some way.

The reflective and absorbing boundary conditions can be defined in a straightforward way by introducing ancillary
states at the boundaries \cite{lehman2012}. The stationary anyons reside only in the bulk, so the braiding
is trivial when the walker moves between the bulk and the boundary states. For periodic boundary conditions the
situation is slightly different, since the transitions between lattice sites $1$ and $N$ must be accompanied by
changes of labels of the fusion charge. For example, if the walker moves from site $N$ to site $1$, the fusion
label $N_A$ changes to $1$ and all other labels are added by one. This operation is represented by the braid word
$\mathcal{B}_{N_A\rightarrow1} = \prod_{k=1}^{N_A-1}b_k$
and the inverse operation is given by $\mathcal{B}_{1\rightarrow N_A}=\mathcal{B}_{N_A\rightarrow1}^\dag$.

The anyonic quantum walk on a finite chain can be simulated numerically by constructing the initial state
$\ket{\Psi(0)}=\ket{s_0}\ket{\Phi_0}\ket{c_0}$ and applying the time evolution operator $t$ times,
$\ket{\Psi(t)}=W^t\ket{\Psi(0)}$. The probability distribution at time step $t$ is given by the diagonal values
of the reduced density matrix of the position space, see Eq. (\ref{eq:aqwprobdistr1}),
and the braid generators $b_s$ for Ising and $SU(2)_k$ anyons can be constructed using Eq. (\ref{eq:braidgens}).
It is known that on finite graphs, the quantum walk does not converge into any asymptotic distribution
(unless the system is initially in an eigenstate of the time evolution operator, in which case the walk is trivial).
However, the time-averaged probability distribution,
\begin{equation}
\overline{p(s,t)} = \frac{1}{t+1}\sum_{\tau=0}^tp(s,\tau)
\end{equation}
has been shown to converge \cite{aakv2001}, but the asymptotic distribution is not necessarily the
uniform distribution.
As the classical random walk is a Markov process, any information on the initial state is lost after a certain
number of time steps, and the probability distribution becomes uniform on the nodes in the infinite time limit.
One can therefore compare the convergence of the random walk and the time-averaged quantum walk to their
respective asymptotic distributions.

\subsection{Mixing time}

The convergence to the asymptotic distribution can be quantified by different measures.
One such measure is the $\epsilon$-mixing time defined as 
\begin{equation}
M_\epsilon = \min\{T|\; ||D_t-\pi||\leq\epsilon,\;\forall t\geq T\}
\end{equation}
where $D_t$ is the probability distribution at time step $t$, the asymptotic distribution is
$\pi=\lim\limits_{t\rightarrow\infty}D_t$, and the distance measure between two probability distributions is
the total variation distance defined as $||D_1-D_2||=\sum_i|D_1(i)-D_2(i)|$.
The mixing time of random walks is shown to converge as $M_\epsilon\sim O(N^2\log1/\epsilon)$.
Aharonov {\it et al.} showed that the quantum walk converges as $M_\epsilon\sim O(N\log N\:1/\epsilon^3)$,
which provides a quadratic speedup in number of sites $N$, but an exponential slowdown in the parameter
$\epsilon$. The convergence with respect to $\epsilon$ can be improved by amplification, and Richter has
shown that a suitably chosen amplification scheme can enhance the convergence to
$O(N\log1/\epsilon)$ \cite{richter2007}.

\begin{figure}
\begin{center}
\begingroup
  \makeatletter
  \providecommand\color[2][]{%
    \GenericError{(gnuplot) \space\space\space\@spaces}{%
      Package color not loaded in conjunction with
      terminal option `colourtext'%
    }{See the gnuplot documentation for explanation.%
    }{Either use 'blacktext' in gnuplot or load the package
      color.sty in LaTeX.}%
    \renewcommand\color[2][]{}%
  }%
  \providecommand\includegraphics[2][]{%
    \GenericError{(gnuplot) \space\space\space\@spaces}{%
      Package graphicx or graphics not loaded%
    }{See the gnuplot documentation for explanation.%
    }{The gnuplot epslatex terminal needs graphicx.sty or graphics.sty.}%
    \renewcommand\includegraphics[2][]{}%
  }%
  \providecommand\rotatebox[2]{#2}%
  \@ifundefined{ifGPcolor}{%
    \newif\ifGPcolor
    \GPcolortrue
  }{}%
  \@ifundefined{ifGPblacktext}{%
    \newif\ifGPblacktext
    \GPblacktexttrue
  }{}%
  \let\gplgaddtomacro\g@addto@macro
  \gdef\gplbacktext{}%
  \gdef\gplfronttext{}%
  \makeatother
  \ifGPblacktext
    \def\colorrgb#1{}%
    \def\colorgray#1{}%
  \else
    \ifGPcolor
      \def\colorrgb#1{\color[rgb]{#1}}%
      \def\colorgray#1{\color[gray]{#1}}%
      \expandafter\def\csname LTw\endcsname{\color{white}}%
      \expandafter\def\csname LTb\endcsname{\color{black}}%
      \expandafter\def\csname LTa\endcsname{\color{black}}%
      \expandafter\def\csname LT0\endcsname{\color[rgb]{1,0,0}}%
      \expandafter\def\csname LT1\endcsname{\color[rgb]{0,1,0}}%
      \expandafter\def\csname LT2\endcsname{\color[rgb]{0,0,1}}%
      \expandafter\def\csname LT3\endcsname{\color[rgb]{1,0,1}}%
      \expandafter\def\csname LT4\endcsname{\color[rgb]{0,1,1}}%
      \expandafter\def\csname LT5\endcsname{\color[rgb]{1,1,0}}%
      \expandafter\def\csname LT6\endcsname{\color[rgb]{0,0,0}}%
      \expandafter\def\csname LT7\endcsname{\color[rgb]{1,0.3,0}}%
      \expandafter\def\csname LT8\endcsname{\color[rgb]{0.5,0.5,0.5}}%
    \else
      \def\colorrgb#1{\color{black}}%
      \def\colorgray#1{\color[gray]{#1}}%
      \expandafter\def\csname LTw\endcsname{\color{white}}%
      \expandafter\def\csname LTb\endcsname{\color{black}}%
      \expandafter\def\csname LTa\endcsname{\color{black}}%
      \expandafter\def\csname LT0\endcsname{\color{black}}%
      \expandafter\def\csname LT1\endcsname{\color{black}}%
      \expandafter\def\csname LT2\endcsname{\color{black}}%
      \expandafter\def\csname LT3\endcsname{\color{black}}%
      \expandafter\def\csname LT4\endcsname{\color{black}}%
      \expandafter\def\csname LT5\endcsname{\color{black}}%
      \expandafter\def\csname LT6\endcsname{\color{black}}%
      \expandafter\def\csname LT7\endcsname{\color{black}}%
      \expandafter\def\csname LT8\endcsname{\color{black}}%
    \fi
  \fi
  \setlength{\unitlength}{0.0500bp}%
  \begin{picture}(4104.00,2872.00)%
    \gplgaddtomacro\gplbacktext{%
      \csname LTb\endcsname%
      \put(924,660){\makebox(0,0)[r]{\strut{} 0}}%
      \put(924,1042){\makebox(0,0)[r]{\strut{} 0.2}}%
      \put(924,1424){\makebox(0,0)[r]{\strut{} 0.4}}%
      \put(924,1806){\makebox(0,0)[r]{\strut{} 0.6}}%
      \put(924,2188){\makebox(0,0)[r]{\strut{} 0.8}}%
      \put(924,2570){\makebox(0,0)[r]{\strut{} 1}}%
      \put(1056,440){\makebox(0,0){\strut{} 0}}%
      \put(1553,440){\makebox(0,0){\strut{} 10}}%
      \put(2050,440){\makebox(0,0){\strut{} 20}}%
      \put(2547,440){\makebox(0,0){\strut{} 30}}%
      \put(3043,440){\makebox(0,0){\strut{} 40}}%
      \put(3540,440){\makebox(0,0){\strut{} 50}}%
      \put(4037,440){\makebox(0,0){\strut{} 60}}%
      \put(154,1710){\rotatebox{-270}{\makebox(0,0){\strut{}$D(t,1000)$}}}%
      \put(2546,110){\makebox(0,0){\strut{}Time step $t$}}%
      \put(0,2727){\makebox(0,0)[l]{\strut{}\textbf{a)}}}%
    }%
    \gplgaddtomacro\gplfronttext{%
      \csname LTb\endcsname%
      \put(3182,2441){\makebox(0,0)[r]{\strut{}\small RW}}%
      \csname LTb\endcsname%
      \put(3182,2221){\makebox(0,0)[r]{\strut{}RW$_{\text{avg}}$}}%
      \csname LTb\endcsname%
      \put(3182,2001){\makebox(0,0)[r]{\strut{}QW$_{\text{avg}}$}}%
      \csname LTb\endcsname%
      \put(3182,1781){\makebox(0,0)[r]{\strut{}Ising$_{\text{avg}}$}}%
    }%
    \gplbacktext
    \put(0,0){\includegraphics{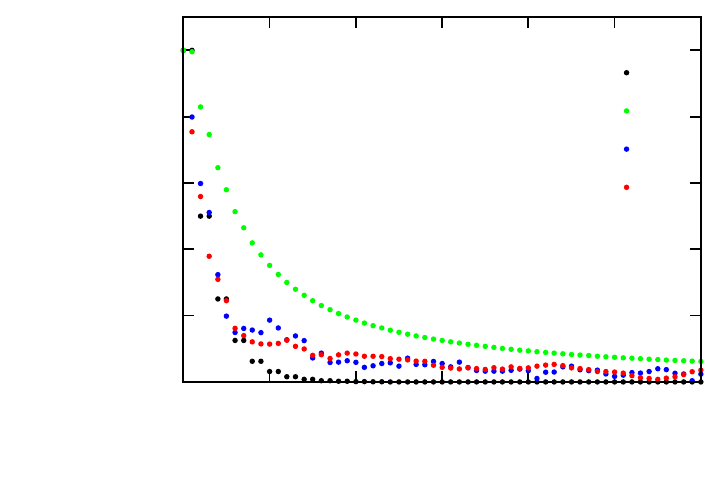}}%
    \gplfronttext
  \end{picture}%
\endgroup
\begingroup
  \makeatletter
  \providecommand\color[2][]{%
    \GenericError{(gnuplot) \space\space\space\@spaces}{%
      Package color not loaded in conjunction with
      terminal option `colourtext'%
    }{See the gnuplot documentation for explanation.%
    }{Either use 'blacktext' in gnuplot or load the package
      color.sty in LaTeX.}%
    \renewcommand\color[2][]{}%
  }%
  \providecommand\includegraphics[2][]{%
    \GenericError{(gnuplot) \space\space\space\@spaces}{%
      Package graphicx or graphics not loaded%
    }{See the gnuplot documentation for explanation.%
    }{The gnuplot epslatex terminal needs graphicx.sty or graphics.sty.}%
    \renewcommand\includegraphics[2][]{}%
  }%
  \providecommand\rotatebox[2]{#2}%
  \@ifundefined{ifGPcolor}{%
    \newif\ifGPcolor
    \GPcolortrue
  }{}%
  \@ifundefined{ifGPblacktext}{%
    \newif\ifGPblacktext
    \GPblacktexttrue
  }{}%
  \let\gplgaddtomacro\g@addto@macro
  \gdef\gplbacktext{}%
  \gdef\gplfronttext{}%
  \makeatother
  \ifGPblacktext
    \def\colorrgb#1{}%
    \def\colorgray#1{}%
  \else
    \ifGPcolor
      \def\colorrgb#1{\color[rgb]{#1}}%
      \def\colorgray#1{\color[gray]{#1}}%
      \expandafter\def\csname LTw\endcsname{\color{white}}%
      \expandafter\def\csname LTb\endcsname{\color{black}}%
      \expandafter\def\csname LTa\endcsname{\color{black}}%
      \expandafter\def\csname LT0\endcsname{\color[rgb]{1,0,0}}%
      \expandafter\def\csname LT1\endcsname{\color[rgb]{0,1,0}}%
      \expandafter\def\csname LT2\endcsname{\color[rgb]{0,0,1}}%
      \expandafter\def\csname LT3\endcsname{\color[rgb]{1,0,1}}%
      \expandafter\def\csname LT4\endcsname{\color[rgb]{0,1,1}}%
      \expandafter\def\csname LT5\endcsname{\color[rgb]{1,1,0}}%
      \expandafter\def\csname LT6\endcsname{\color[rgb]{0,0,0}}%
      \expandafter\def\csname LT7\endcsname{\color[rgb]{1,0.3,0}}%
      \expandafter\def\csname LT8\endcsname{\color[rgb]{0.5,0.5,0.5}}%
    \else
      \def\colorrgb#1{\color{black}}%
      \def\colorgray#1{\color[gray]{#1}}%
      \expandafter\def\csname LTw\endcsname{\color{white}}%
      \expandafter\def\csname LTb\endcsname{\color{black}}%
      \expandafter\def\csname LTa\endcsname{\color{black}}%
      \expandafter\def\csname LT0\endcsname{\color{black}}%
      \expandafter\def\csname LT1\endcsname{\color{black}}%
      \expandafter\def\csname LT2\endcsname{\color{black}}%
      \expandafter\def\csname LT3\endcsname{\color{black}}%
      \expandafter\def\csname LT4\endcsname{\color{black}}%
      \expandafter\def\csname LT5\endcsname{\color{black}}%
      \expandafter\def\csname LT6\endcsname{\color{black}}%
      \expandafter\def\csname LT7\endcsname{\color{black}}%
      \expandafter\def\csname LT8\endcsname{\color{black}}%
    \fi
  \fi
  \setlength{\unitlength}{0.0500bp}%
  \begin{picture}(4104.00,2872.00)%
    \gplgaddtomacro\gplbacktext{%
      \csname LTb\endcsname%
      \put(924,660){\makebox(0,0)[r]{\strut{} 0}}%
      \put(924,1042){\makebox(0,0)[r]{\strut{} 0.2}}%
      \put(924,1424){\makebox(0,0)[r]{\strut{} 0.4}}%
      \put(924,1806){\makebox(0,0)[r]{\strut{} 0.6}}%
      \put(924,2188){\makebox(0,0)[r]{\strut{} 0.8}}%
      \put(924,2570){\makebox(0,0)[r]{\strut{} 1}}%
      \put(1056,440){\makebox(0,0){\strut{} 0}}%
      \put(1652,440){\makebox(0,0){\strut{} 100}}%
      \put(2248,440){\makebox(0,0){\strut{} 200}}%
      \put(2845,440){\makebox(0,0){\strut{} 300}}%
      \put(3441,440){\makebox(0,0){\strut{} 400}}%
      \put(4037,440){\makebox(0,0){\strut{} 500}}%
      \put(2546,110){\makebox(0,0){\strut{}Time step $t$}}%
      \put(205,2727){\makebox(0,0)[l]{\strut{}\textbf{b)}}}%
    }%
    \gplgaddtomacro\gplfronttext{%
    }%
    \gplbacktext
    \put(0,0){\includegraphics{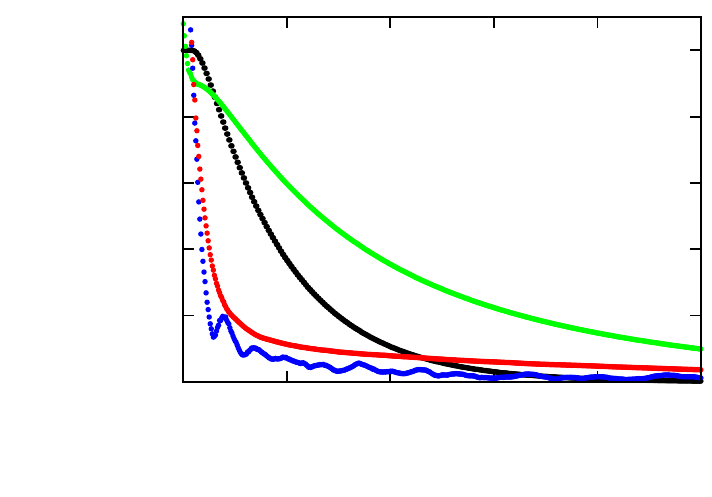}}%
    \gplfronttext
  \end{picture}%
\endgroup
\end{center}
\caption{Total variation distance $D(t,T)$ between probability distributions at time steps $t$ and $T$.
a) 5 sites. b) 21 sites. Black represents the classical random walk
and green represents the time-averaged classical random walk. Blue is the time-averaged Hadamard walk and
red is the time-averaged Ising walk.
Total number of time steps was 1000, periodic boundary conditions were used, and the initial fusion state $\ket{\Phi_0}$
was chosen such that tracing scheme is plat closure.
}
\label{fig:mixingdistance}
\end{figure}

As the asymptotic distribution is not currently known for anyonic walks,
an operational measure of convergence is defined as the total
variation distance to the final distribution at time step $T$:
\begin{equation}
D(t,T) = \big|\big|\overline{p(s,t)}-\overline{p(s,T)}\big|\big|
\end{equation}
such that lower values of $D$ mean faster convergence. The total variation distance for Ising anyons is plotted
in Fig. \ref{fig:mixingdistance} with the corresponding plots of the random walk and the Hadamard walk
(also the average distribution of the random walk is plotted for comparison).
For a small number of sites, the Ising walk converges in a similar fashion as the Hadamard walk. The total variation
distance is initially similar to the classical random walk (RW), and fluctuates between RW and RW$_{\text{avg}}$ at
later time steps (mixing time for RW is calculated using the instantaneous probability distribution and RW$_{\text{avg}}$
is calculated using the time-averaged distribution, as with quantum walks).
For a larger number of sites, the convergence of the Ising walk is still similar to the Hadamard walk.
Both walks converge faster than the RW initially, but in the course of time the total variation distance
of both the Ising and Hadamard walks settles between RW and RW$_{\text{avg}}$.
For a large number of sites, the convergence of the Ising walk is quite smooth, whereas the convergence of the
Hadamard walk is not uniform.

It is curious to note that the mixing properties of the non-Abelian anyonic walk are quite similar to those of the
quantum walk. This is in contrast to the results on infinite chains as discussed in the previous section,
where the propagation of the non-Abelian anyon shows similar properties to the classical random walk.
Analysis of the probability distributions in finite chains shows that both the anyonic and non-anyonic quantum
walks evolve in a disordered manner, and the probability distributions between individual time steps can be
very different, whereas the probability distributions of the classical random walk are similar.
The Ising and Hadamard walks sample from a wider range of probability distributions, and therefore the
average distribution converges slower than RW. However, the Ising and Hadamard walks converge faster than
RW$_{\text{avg}}$.

\section{Proposal for experimental setup} \label{sec:experimental}

A possible experimental platform for simulating anyonic walks could be in Fractional Quantum Hall samples.
The quasiparticles at the $\nu=5/2$ plateau have been proposed to have non-Abelian anyonic statistics corresponding
to Ising anyons.
The experiments probing quasiparticle statistics can be done using a Fabry-Perot type interferometer
on top of a FQHE layer \cite{cfksw1997,fntw1998}. An interference pattern is induced on the longitudinal
conductivity $\sigma_{xx}$ as the side gate voltage $V_s$ is varied.

The statistical phases of anyons are generally very hard to detect. The interference pattern of $\sigma_{xx}$
is dominated by Aharonov-B\"ohm oscillations, the period of which depends on the effective electric charge $e^*$ of
the quasiparticles in the medium. The effective charge of quasiparticles at the $\nu=5/2$ plateau is $e^*=e/4$
in all proposed models, regardless of whether the statistics is Abelian or non-Abelian, so the period of
Aharonov-B\"ohm oscillations does not distinguish between the statistics. It was proposed however that if the
quasiparticles are non-Abelian and if there is an odd number of them between the tunneling point contacts
of the interferometer, the Aharonov-B\"ohm oscillations disappear \cite{sh2006,bks2006}.
The disappearance of the oscillations was confirmed experimentally by Willett \textit{et al} \cite{wpw},
but instead of complete suppression of the oscillations there seem to be
persistent $e/2$ oscillations when the number of quasiparticles is odd. The $e/2$ oscillations could result from
tunneling of Abelian quasiparticles with effective charge $e^*=e/2$, or non-Abelian quasiparticles tunnelling
more than once. Analysis by Bishara \textit{et al.} shows that the first option is more plausible \cite{bbnss2009}.

\subsection{Setup for measurements}

We propose that an experimental setup which implements the anyonic quantum walk scheme could also be used to probe
the statistical properties of quasiparticles.
Such a scheme relies on transport measurements of edge mode quasiparticles in a multipoint contact version of the
Fabry-Perot type interferometer \cite{lzbpw2011}. Instead of measuring the interference pattern of conductivity when the side
gate voltage is varied, the conductivity could be measured in equilibrium while keeping the side gate voltage
constant. We suggest that the quasiparticle statistics has an experimental signature on the
conductivity along the point contact array, and that this signature becomes clearer when the
number of tunneling point contacts increases.

To measure the probability distribution of the quantum walk directly would require introducing
contacts at each location corresponding to spatial sites of the quantum walk.
In a simplified scheme, very similar to that used in two-point contact interferometry, the conductivity would only measured at the ends of the multipoint contact ladder.
Quasiparticles are ejected to the array from the middle or from the other end via an external contact.
Changing the voltage by the chemical potential of the quasiparticles creates pairs of quasiparticles
on the edges.
Thus, their wavepackets are initially localized at the location of the contact, and they start to
perform a quantum walk across the ladder.
To measure the conductivity between two points, one has to introduce another contact at some other point
in the ladder. If the measurement destroys the quasiparticle, this corresponds to introducing an absorbing
boundary to the quantum walk.
The conductivity between the ejection point and the measurement point is proportional
to the probability to reach the absorbing boundary when the walker starts localized at the ejection point.

\subsection{Exit probability}

\begin{figure}
\begin{center}
\begingroup
  \makeatletter
  \providecommand\color[2][]{%
    \GenericError{(gnuplot) \space\space\space\@spaces}{%
      Package color not loaded in conjunction with
      terminal option `colourtext'%
    }{See the gnuplot documentation for explanation.%
    }{Either use 'blacktext' in gnuplot or load the package
      color.sty in LaTeX.}%
    \renewcommand\color[2][]{}%
  }%
  \providecommand\includegraphics[2][]{%
    \GenericError{(gnuplot) \space\space\space\@spaces}{%
      Package graphicx or graphics not loaded%
    }{See the gnuplot documentation for explanation.%
    }{The gnuplot epslatex terminal needs graphicx.sty or graphics.sty.}%
    \renewcommand\includegraphics[2][]{}%
  }%
  \providecommand\rotatebox[2]{#2}%
  \@ifundefined{ifGPcolor}{%
    \newif\ifGPcolor
    \GPcolortrue
  }{}%
  \@ifundefined{ifGPblacktext}{%
    \newif\ifGPblacktext
    \GPblacktexttrue
  }{}%
  \let\gplgaddtomacro\g@addto@macro
  \gdef\gplbacktext{}%
  \gdef\gplfronttext{}%
  \makeatother
  \ifGPblacktext
    \def\colorrgb#1{}%
    \def\colorgray#1{}%
  \else
    \ifGPcolor
      \def\colorrgb#1{\color[rgb]{#1}}%
      \def\colorgray#1{\color[gray]{#1}}%
      \expandafter\def\csname LTw\endcsname{\color{white}}%
      \expandafter\def\csname LTb\endcsname{\color{black}}%
      \expandafter\def\csname LTa\endcsname{\color{black}}%
      \expandafter\def\csname LT0\endcsname{\color[rgb]{1,0,0}}%
      \expandafter\def\csname LT1\endcsname{\color[rgb]{0,1,0}}%
      \expandafter\def\csname LT2\endcsname{\color[rgb]{0,0,1}}%
      \expandafter\def\csname LT3\endcsname{\color[rgb]{1,0,1}}%
      \expandafter\def\csname LT4\endcsname{\color[rgb]{0,1,1}}%
      \expandafter\def\csname LT5\endcsname{\color[rgb]{1,1,0}}%
      \expandafter\def\csname LT6\endcsname{\color[rgb]{0,0,0}}%
      \expandafter\def\csname LT7\endcsname{\color[rgb]{1,0.3,0}}%
      \expandafter\def\csname LT8\endcsname{\color[rgb]{0.5,0.5,0.5}}%
    \else
      \def\colorrgb#1{\color{black}}%
      \def\colorgray#1{\color[gray]{#1}}%
      \expandafter\def\csname LTw\endcsname{\color{white}}%
      \expandafter\def\csname LTb\endcsname{\color{black}}%
      \expandafter\def\csname LTa\endcsname{\color{black}}%
      \expandafter\def\csname LT0\endcsname{\color{black}}%
      \expandafter\def\csname LT1\endcsname{\color{black}}%
      \expandafter\def\csname LT2\endcsname{\color{black}}%
      \expandafter\def\csname LT3\endcsname{\color{black}}%
      \expandafter\def\csname LT4\endcsname{\color{black}}%
      \expandafter\def\csname LT5\endcsname{\color{black}}%
      \expandafter\def\csname LT6\endcsname{\color{black}}%
      \expandafter\def\csname LT7\endcsname{\color{black}}%
      \expandafter\def\csname LT8\endcsname{\color{black}}%
    \fi
  \fi
  \setlength{\unitlength}{0.0500bp}%
  \begin{picture}(4104.00,2872.00)%
    \gplgaddtomacro\gplbacktext{%
      \csname LTb\endcsname%
      \put(924,660){\makebox(0,0)[r]{\strut{} 0}}%
      \put(924,1049){\makebox(0,0)[r]{\strut{} 0.2}}%
      \put(924,1439){\makebox(0,0)[r]{\strut{} 0.4}}%
      \put(924,1828){\makebox(0,0)[r]{\strut{} 0.6}}%
      \put(924,2218){\makebox(0,0)[r]{\strut{} 0.8}}%
      \put(924,2607){\makebox(0,0)[r]{\strut{} 1}}%
      \put(1056,440){\makebox(0,0){\strut{} 0}}%
      \put(1531,440){\makebox(0,0){\strut{} 50}}%
      \put(2006,440){\makebox(0,0){\strut{} 100}}%
      \put(2481,440){\makebox(0,0){\strut{} 150}}%
      \put(2955,440){\makebox(0,0){\strut{} 200}}%
      \put(3430,440){\makebox(0,0){\strut{} 250}}%
      \put(3905,440){\makebox(0,0){\strut{} 300}}%
      \put(154,1633){\rotatebox{-270}{\makebox(0,0){\strut{}$P_{\text{ex}}(t)$}}}%
      \put(2480,110){\makebox(0,0){\strut{}Time step $t$}}%
      \put(0,2727){\makebox(0,0)[l]{\strut{}\textbf{a)}}}%
    }%
    \gplgaddtomacro\gplfronttext{%
      \csname LTb\endcsname%
      \put(3192,1524){\makebox(0,0)[r]{\strut{}RW}}%
      \csname LTb\endcsname%
      \put(3192,1304){\makebox(0,0)[r]{\strut{}QW}}%
      \csname LTb\endcsname%
      \put(3192,1084){\makebox(0,0)[r]{\strut{}Ising}}%
    }%
    \gplbacktext
    \put(0,0){\includegraphics{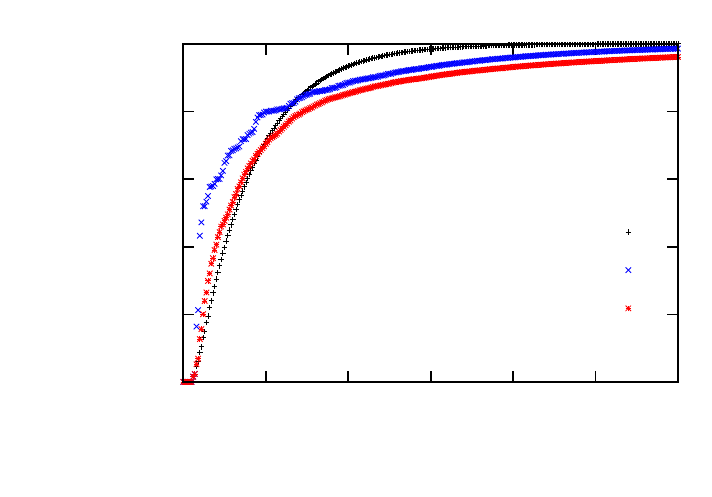}}%
    \gplfronttext
  \end{picture}%
\endgroup
\hspace{2em}
\begingroup
  \makeatletter
  \providecommand\color[2][]{%
    \GenericError{(gnuplot) \space\space\space\@spaces}{%
      Package color not loaded in conjunction with
      terminal option `colourtext'%
    }{See the gnuplot documentation for explanation.%
    }{Either use 'blacktext' in gnuplot or load the package
      color.sty in LaTeX.}%
    \renewcommand\color[2][]{}%
  }%
  \providecommand\includegraphics[2][]{%
    \GenericError{(gnuplot) \space\space\space\@spaces}{%
      Package graphicx or graphics not loaded%
    }{See the gnuplot documentation for explanation.%
    }{The gnuplot epslatex terminal needs graphicx.sty or graphics.sty.}%
    \renewcommand\includegraphics[2][]{}%
  }%
  \providecommand\rotatebox[2]{#2}%
  \@ifundefined{ifGPcolor}{%
    \newif\ifGPcolor
    \GPcolortrue
  }{}%
  \@ifundefined{ifGPblacktext}{%
    \newif\ifGPblacktext
    \GPblacktexttrue
  }{}%
  \let\gplgaddtomacro\g@addto@macro
  \gdef\gplbacktext{}%
  \gdef\gplfronttext{}%
  \makeatother
  \ifGPblacktext
    \def\colorrgb#1{}%
    \def\colorgray#1{}%
  \else
    \ifGPcolor
      \def\colorrgb#1{\color[rgb]{#1}}%
      \def\colorgray#1{\color[gray]{#1}}%
      \expandafter\def\csname LTw\endcsname{\color{white}}%
      \expandafter\def\csname LTb\endcsname{\color{black}}%
      \expandafter\def\csname LTa\endcsname{\color{black}}%
      \expandafter\def\csname LT0\endcsname{\color[rgb]{1,0,0}}%
      \expandafter\def\csname LT1\endcsname{\color[rgb]{0,1,0}}%
      \expandafter\def\csname LT2\endcsname{\color[rgb]{0,0,1}}%
      \expandafter\def\csname LT3\endcsname{\color[rgb]{1,0,1}}%
      \expandafter\def\csname LT4\endcsname{\color[rgb]{0,1,1}}%
      \expandafter\def\csname LT5\endcsname{\color[rgb]{1,1,0}}%
      \expandafter\def\csname LT6\endcsname{\color[rgb]{0,0,0}}%
      \expandafter\def\csname LT7\endcsname{\color[rgb]{1,0.3,0}}%
      \expandafter\def\csname LT8\endcsname{\color[rgb]{0.5,0.5,0.5}}%
    \else
      \def\colorrgb#1{\color{black}}%
      \def\colorgray#1{\color[gray]{#1}}%
      \expandafter\def\csname LTw\endcsname{\color{white}}%
      \expandafter\def\csname LTb\endcsname{\color{black}}%
      \expandafter\def\csname LTa\endcsname{\color{black}}%
      \expandafter\def\csname LT0\endcsname{\color{black}}%
      \expandafter\def\csname LT1\endcsname{\color{black}}%
      \expandafter\def\csname LT2\endcsname{\color{black}}%
      \expandafter\def\csname LT3\endcsname{\color{black}}%
      \expandafter\def\csname LT4\endcsname{\color{black}}%
      \expandafter\def\csname LT5\endcsname{\color{black}}%
      \expandafter\def\csname LT6\endcsname{\color{black}}%
      \expandafter\def\csname LT7\endcsname{\color{black}}%
      \expandafter\def\csname LT8\endcsname{\color{black}}%
    \fi
  \fi
  \setlength{\unitlength}{0.0500bp}%
  \begin{picture}(4104.00,2872.00)%
    \gplgaddtomacro\gplbacktext{%
      \csname LTb\endcsname%
      \put(924,660){\makebox(0,0)[r]{\strut{} 0}}%
      \put(924,1049){\makebox(0,0)[r]{\strut{} 0.2}}%
      \put(924,1439){\makebox(0,0)[r]{\strut{} 0.4}}%
      \put(924,1828){\makebox(0,0)[r]{\strut{} 0.6}}%
      \put(924,2218){\makebox(0,0)[r]{\strut{} 0.8}}%
      \put(924,2607){\makebox(0,0)[r]{\strut{} 1}}%
      \put(1056,440){\makebox(0,0){\strut{} 0}}%
      \put(1626,440){\makebox(0,0){\strut{} 200}}%
      \put(2196,440){\makebox(0,0){\strut{} 400}}%
      \put(2765,440){\makebox(0,0){\strut{} 600}}%
      \put(3335,440){\makebox(0,0){\strut{} 800}}%
      \put(3905,440){\makebox(0,0){\strut{} 1000}}%
      \put(2480,110){\makebox(0,0){\strut{}Time step $t$}}%
      \put(0,2727){\makebox(0,0)[l]{\strut{}\textbf{b)}}}%
    }%
    \gplgaddtomacro\gplfronttext{%
    }%
    \gplbacktext
    \put(0,0){\includegraphics{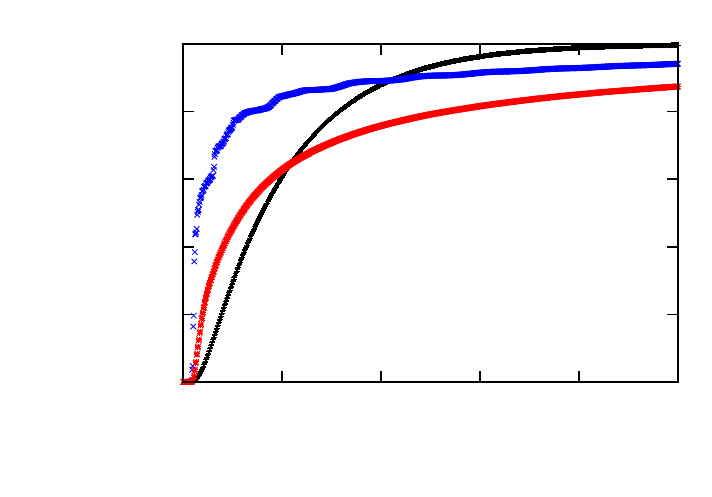}}%
    \gplfronttext
  \end{picture}%
\endgroup
\end{center}
\caption{Accumulated exit probability as a function of time. The number of sites is
a) $N=12$ and b) $N=28$.}
\label{fig:exitprob}
\end{figure}

To illustrate how statistics can affect the observed conductivity,
consider a scheme where quasiparticles are ejected from the middle of the ladder,
and the sum of conductivities at the ends of the ladder is measured.
In principle the conductivity is related to the probability to
propagate from the initial site to the edges of the array. Starting from the initial site $s_0$
at {$t$}{$=$}{$0$}, this probability is called the accumulated exit probability defined by
\begin{equation}
P_{\text{ex}}(t) = \sum_{\tau=0}^t p(0,\tau)+p(N+1,\tau),
\end{equation}
where the sites $s=\{0,N+1\}$ are absorbing boundary sites.
The exit probability is plotted in Fig. \ref{fig:exitprob} as a function of time for the classical
random walk, the Hadamard walk and the Ising anyon walk.
At short time scales the Ising walk and RW behave quite similarly. The probability to observe the
particle is higher for QW, as is expected for ballistic propagation of the walker.
At long time scales the probability to observe the RW particle is the highest and it approaches $1$.
The probability to observe the Ising anyon is the lowest. Both the Ising walk and the QW
probabilities seem to approach a constant $<1$, which is in agreement with earlier results which
show that the asymptotic exit probability does not approach unity \cite{abnvw2001}.
It should also be noted that these effects become clearer when the number of sites $N$ grows.

A few problems with these proposals can be immediately identified. The results presented here
use the Hadamard coin which corresponds to strong backscattering limit at the quantum point
contacts, which is known to be problematic in observation of statistics.  However, the main results shouldn't be too sensitive to the choice of coin because even for small coin rotation angles, there will entanglement generated for non-Abelian anyons which will affect the transport behavior in a similar manner though over a longer time scale. Also, it might be
hard to protect the mobile quasiparticles from spatial decoherence, such that they would propagate
coherently across several point contacts across the sample.  If this were the case then Abelian and non-Abelian anyons would give similar signatures.  Finally, the quasiparticle occupations
on the islands and gate voltages might fluctuate between experimental runs.
In this case the model would have to take the effects of disorder into account \cite{zlspb2012}.

\section{Discussion and outlook}

The effects of particle statistics in the one-dimensional anyonic quantum walk were analysed,
and the results on the infinite line indicate that the walker propagates diffusively, showing
similarities with classical random walks and quantum walks with decoherence. In finite chains,
the effects are less pronounced. The mixing properties of the average probability distribution 
are close to those of the Hadamard quantum walk, and the probability to be absorbed at the
boundaries is between those of RW and QW. Experimental realizations using an array of quantum point
contacts was also discussed.

One interesting generalization of the models presented here is obtained by considering non-uniform
charge configurations \cite{zlspb2012}. Some work is also being done with continuous-time anyonic
quantum walks using matrix product states and Hubbard-type Hamiltonian models.
Other natural generalizations include multiparticle walks with more than one mobile anyon and
anyonic quantum walks in two-dimensional lattices.

\end{document}